\documentclass[twocolumn,showpacs,preprintnumbers,amsmath,amssymb]{revtex4}
\usepackage{graphicx}
\usepackage{bm}

\def\NPA{{Nucl. Phys.} {\bf A}}
\def\NPB{{Nucl. Phys.} {\bf B}}
\def\PLB{{Phys. Lett.} B}

\def\PRD{{Phys. Rev.} D}
\def\ZPA{{Z. Phys.} A}
\begin{document}
\newcommand{\ttbs}{\char'134}


\title{What the Gribov copy tells on the confinement and\\
the theory of dynamical chiral symmetry breaking}
\author{Sadataka Furui}
\email{furui@umb.teikyo-u.ac.jp}
\homepage{http://albert.umb.teikyo-u.ac.jp/furui_lab/furuipbs.htm}
\affiliation{%
School of Science and Engineering, Teikyo University, 320-8551 Japan.
}%
\author{Hideo Nakajima}
\email{nakajima@is.utsunomiya-u.ac.jp}
 
\affiliation{
Department of Information Science, Utsunomiya University, 321-8585 Japan. 
}%

\date{\today}

\begin{abstract}
  We performed lattice Landau gauge QCD simulation on $\beta=6.0, 16^4, 24^4, 32^4$ and $\beta=6.4, 32^4, 48^4$ and $56^4$ by adopting the gauge fixing that minimizes the norm of the gauge field, and measured the running coupling by using the gluon propagator and the ghost propagator. In view of ambiguity in the vertex renormalization factor $\tilde Z_1$ in the lattice, we adjust the normalization of the running coupling by the perturbative QCD results near the highest momentum point. It has a maximum  $\alpha_s(q)\simeq 2.1(3)$ at around $q=0.5$ GeV and decreases as $q$ approaches 0, and the Kugo-Ojima parameter reached -0.83(2). The infrared exponent of the ghost propagator at 0.4GeV region is $\alpha_G=0.20$  but there is an exceptional Gribov copy with $\alpha_G=0.27$.
The features of the exceptional Gribov copy are investigated by measuring four one-dimensional Fourier transform(1-d FT) of the gluon propagator transverse to each lattice axis. We observe, in general, correlation between absolute value of the Kugo-Ojima parameter and the degree of reflection positivity violation in the 1-d FT of the gluon propagator.@The 1-d FT of the exceptional Gribov copy has an axis whose sample-wise gluon propagator manifestly violates reflection positivity, and the average of the Cartan subalgebra components of the Kugo-Ojima parameter along this axis is consistent to -1. The running coupling of the ensemble average shows a suppression at 0 momentum, but when the ghost propagator of the exceptional Gribov copy is adopted, the suppression disappears and the data implies presence of the infrared fixed point $\alpha_s(0)\sim 2.5(5)$ and $\kappa=0.5$ suggested by the Dyson-Schwinger approach in the multiplicative renormalizable scheme. Comparison with the SU(2) QCD and $N_f=2$ unquenched SU(3) QCD are also made.

\end{abstract}

\maketitle

\section{\label{sec:level1}Introduction}
The lattice Landau gauge QCD simulation suffers from Gribov copy problem and
its effect on the confinement was discussed by several authors \cite{Gr, Zw, Zw1, Nah}. As a method for obtaining the unique gauge, we adopted the fundamental modular gauge (FMG) i.e. a configuration with the minimum norm of the gauge field and studied the Gribov copy problem in SU(2) \cite{lat03}. We compared the absolute minimum  configuration obtained by the Landau gauge fixing via the parallel tempering method and the 1st copy  which is obtained by our straightforward Landau gauge fixing.
We observed that the FMG configurations and the 1st copy which is in the Gribov region but not necessarily in the FM region have the following differences: 1) The absolute value of the Kugo-Ojima parameter $c$ \cite{KO,FN03}, which gives the sufficient condition of the confinement,  of the FMG is smaller than that of the 1st copy. 2) The singularity of the ghost propagator of the FMG is less than that of the 1st copy. 3) The gluon propagator of the two copies are almost the same within statistical errors. 4) The horizon function deviation parameter $h$ of the FMG is not closer to 0, i.e. the value expected in the continuum limit, than that of the 1st copy. 

The proximity of the FMG configuration and the boundary of the Gribov region in SU(2)
in $8^4, 12^4$ and $16^4$ lattices with $\beta=0, 0.8, 1.6$ and $2.7$ was studied  in \cite{Cucc}. The tendency that the smallest eigenvalue of the Faddeev-Popov matrix of the FMG and that of the 1st copy come closer as $\beta$ and lattice size become larger was observed, although as remarked in \cite{Cucc} the physical volume of $\beta=2.7$, $16^4$ lattice is small and not close to the continuum limit. Qualitatve features of the profile of the Morse function
\begin{equation}
{\cal E}[g]=\frac{1}{2}\sum_{\mu,a}\int d^4x\{[A_\mu^{(g)}]^a(x)\}^2
\end{equation}
where $g=e^{\epsilon\cdot \lambda}$, 
was sketched as a function with respect to the magnitude of the infinitesimal gauge transformation parameter $\epsilon$ and  a parameter 
 $r$ which is defined by 2nd, 3rd and 4th derivative of ${\cal E}[g]$ with respect to $\epsilon$ at the origin.  The simulation suggests that as the $\beta$ and lattice size become large, the parameter $r$ decreases. The meaning of the parameter $r$ is such that larger $r$ than the critical value implies an existence of a smaller local minimum than that of the origin. 

The difference of the 1st copy and the FMG in the $\beta=2.2$, $16^4$ lattice \cite{lat03} indicates that the FMG does not overlap with the boundary of the Gribov region in that simulation. In the Langevin formulation of QCD, Zwanziger conjectures that the path integral over the FM region will become equivalent to that over the Gribov region in the continuum \cite{Zw1}. This conjecture is consistent with the view that the boundary of the FMG and that of the Gribov region overlaps and the probability distribution is accumulated in this overlapped region. On the lattice, when $\beta$ and the lattice size is not large enough, distribution of Gribov copies i.e. statistical weight of the copies is crucial for extracting sample averages. 

In the previous paper \cite{FN03}, we measured the QCD running coupling and the Kugo-Ojima parameter in $\beta=6.0, 16^4, 24^4, 32^4$ and $\beta=6.4, 32^4$ and $48^4$. 
The running coupling was found maximum of about 1.1 at around $q=0.5$ GeV, and behaved either approaching constant or even decreasing as $q$ approaches zero, and the Kugo-Ojima parameter was getting larger but staying around $-0.8$ in contrast to the expected value $-1$ in the continuum theory. Thus it is necessary to
perform a larger lattice simulation and to study the dependence of the Gribov copy. We encountered a rather exceptional Gribov copy in $\beta=6.4, 56^4$ which is close to the Gribov boundary and we consider it worthwhile to investigate that sample in some details.   
We analyze those data by comparing with continuum theory like Dyson-Schwinger equation(DSE).

There are extensive reviews on DSE for the Yang Mills theory \cite{RW,AS,Kond,LvS}. The solution of DSE depends on ansatz of momentum truncation and what kind of loop diagrams are included. Two decades ago Mandelstam \cite{Mand} projected the DSE for the gluon propagator by ${\cal P}_{\mu\nu}(q)=\delta_{\mu\nu}-q_\mu q_\nu/q^2$ and without including ghosts, assumed the gluon
wavefunction renormalization factor in the form
\begin{equation}
Z(q^2)=\frac{b}{q^2}+C(q^2)\qquad  b=const.
\end{equation}
Later Brown and Pennington \cite{BP} argued that in order to decouple divergent tadpole contribution, it is more appropriate to project the gluon propagator by
${\cal R}_{\mu\nu}(q)=\delta_{\mu\nu}-4q_\mu q_\nu/q^2$. A careful study of inclusion of ghost loop in this DSE was performed by \cite{SHA}, and they showed 
the infrared QCD running coupling in Landau gauge could be finite. 

The divergent QCD running coupling caused difficulty in the model building of dynamical chiral symmetry breaking \cite{Pag,Hig}. In order to get reasonable
values of the quark condensates, infrared finite QCD running coupling was favored. Recent DSE approach with multiplicative renormalizable(MR) truncation
with infrared finite QCD running coupling \cite{Blo, Blo1} suggests that the confinement and the chiral symmetry breaking can be explained by the unique running coupling. We thus compare the running coupling obtained from our lattice simulation and that used in the DSE and study the dependence on the Gribov copy.

 We produced SU(3) gauge configurations by using the heat-bath method,
performed gauge fixing and analyzed lattice Landau gauge configurations of $\beta=6.4$, $56^4$. The $\beta=6.4, 48^4$ and $56^4$ lattices allow measuring the ghost propagator in the momentum range [0.48,14.6] GeV, and [0.41,14.6] GeV, respectively. 
In the present work, 
the gauge field is defined from the link variables as
$\log U$ type:
$$U_{x,\mu}=e^{A_{x,\mu}},\ A_{x,\mu}^{\dag}=-A_{x,\mu}.$$

The fundamental modular gauge (FMG)\cite{Zw} of lattice size $L$ is specified by the
 global minimum along the gauge orbits, i.e., 

$\Lambda_L=\{U|F_{U}(1)={\rm Min}_gF_{U}(g)\}$, 
$\Lambda_L\subset \Omega_L$, 
where $\Omega_L$ is called the  Gribov region (local minima) and \\
$\Omega_L=\{U|-\partial { D(U)}\ge 0\ ,\ \partial A(U)=0\}.$

Here $F_U(g)$ is defined as
$$\displaystyle F_U(g)=||A^g||^2=\frac{1}{(n^2-1)4V}\sum_{x,\mu}{\rm tr}
 \left({{A^g}_{x,\mu}}^{\dag}A^g_{x,\mu}\right).$$

In the gauge transformation 
\begin{equation}
e^{{A^g}_{x, \mu}}=g_x^\dag e^{A_{x, \mu}}g_{x+\mu},
\end{equation}
where  $g=e^{\epsilon\cdot \lambda}$, the value $\epsilon$ is chosen depending on the maximum norm $|\partial A|_{cr}$ as follows.
\begin{itemize}
\item When $|\partial A|>|\partial A|_{cr}$: $\displaystyle \epsilon_x=\frac{\eta'}{\|\partial A\|}\partial A_x$ ($\eta'\sim 0.05$)

\item When $|\partial A|\leq |\partial A|_{cr}$: $\epsilon=(-\partial_\mu D_\mu(A))^{-1}\eta \partial A$ ($\eta=1 \sim 1.6$) \label{epszero}
\end{itemize}

In the second case, calculation of $(-\partial_\mu D_\mu(A))^{-1}$  
 is performed by Newton's method where the linear equation is solved up to third order of the gauge field, and then the Poisson equation
is solved by the multigrid method \cite{NF,FN,scgt}. The accuracy of the gauge fixed configuration characterized by $\partial A(U)=0$ is
$10^{-4}$ in the maximum norm squared which turned out to be about $10^{-15}$ in the $L_2$ norm squared of the gauge field in contrast to about $10^{-12}$ in $48^4$.

In the calculation of the ghost propagator, i.e. inverse Faddeev-Popov (FP) operator, we adopt 
the conjugate gradient (CG) method, whose accuracy of the
solution in the $q < 0.8$GeV region turned out to be less than 5\% in the maximum norm \cite{lat03,FN03}.

In \cite{FN03}, we analyzed these data using a method inspired by the principle of minimal sensitivity (PMS) and/or the effective charge method \cite{PMS,Gru}, the contour-improved perturbation method \cite{HoMa} and the DSE approach \cite{SHA,AS}. We perform the same analysis to the $56^4$ data.

The infrared behavior of the running coupling is tightly related to the mechanism of the dynamical chiral symmetry breaking\cite{Hig, Blo, FA}. The
lattice data are compared with the theory of dynamical chiral symmetry breaking based on the DSE.

In order to study properties of $\tilde Z_1$ and the infrared features, we extend the $16^4$ SU(2) lattice Landau gauge simulation and compare data of $\beta=2.2, 2.3$ and 2.375. 

In sec. II we show some details of the gauge fixing procedure and show
sample dependence of the gluon propagator, 
Kugo-Ojima parameter and QCD running coupling. In sec. III a brief summary of the DSE as well as the recent exact renormalization group approach(ERGE) are presented. We compare lattice data with results of the theoretical analysis of DSE.
The SU(2) lattice Landau gauge simulation data are summarized in sec. IV.
In order to check qualitative differences between the quenched and unquenched
Landau gauge simulation, we performed an exploratory analysis of the configuration produced by the JLQCD\cite{jlqcd}. The results are shown in sec. V.
 Summary and issues on dynamical chiral symmetry breaking is discussed in sec. VI.

\section{Gribov copy and the $56^4$ lattice data}

The magnitude of $|\partial A|_{cr}$ in the gauge transformation is chosen to be 2.2(copy $A$) or 2(copy $B$). In most cases, gauge fixed configurations are almost the some, but in some cases, different $|\partial A|_{cr}$ produce significantly different copies. 

In order to see the difference of the gluon field of the Gribov copies, we
measured the 4 components of 1-dimensional Fourier transform (1-d FT) of the
sample-wise gluon propagator as follows.  
 We consider the gluon propagator 
\begin{eqnarray}
D_{A,\mu\nu}(q)&=&tr\langle \tilde A_\mu(q)\tilde A_\nu(q)^\dagger \rangle \nonumber\\
&=&(\delta_{\mu\nu}-{q_\mu q_\nu\over q^2})D_A(q^2),
\end{eqnarray}\label{scalarfn}
where $\displaystyle \tilde A_\mu(q)=\frac{1}{\sqrt V}\sum_x e^{-iqx}A_\mu(x)$.
In the data analysis of \ref{scalarfn}, there are some possible choice of $q$.
Here we choose $q$ transverse to $\mu$. Since there are 3 possible choices of $\nu\ne \mu$, we make an average of the three combinations $D_A(q^2)_\mu$ 
\begin{eqnarray}
D_A(q^2)&=&\frac{1}{3}\sum_\mu\sum_{\nu\ne \mu}\frac{1}{3}\langle \tilde A_\mu(q_\nu)\tilde A_\mu(q_\nu)^\dagger\rangle\nonumber\\
&=&\frac{1}{3}\sum_\mu D_A(q^2)_\mu
\end{eqnarray}

When the axis $\nu$ is chosen as $t$ axis, and an average over $\mu$ is taken, it is equivalent to the specific Schwinger function
\begin{equation}
S(t,\vec 0)=\frac{1}{\sqrt L}\sum_{q_0=0}^{L-1} D_A(q_0,\vec 0)e^{2\pi i q_0 t/L}
\end{equation}
where $L$ is the lattice size. 

When the Schwinger function becomes negative, the reflection positivity becomes violated, which means that the gluon is not a physical particle. Violation of positivity is considered as a sufficient condition of the confinement\cite{FA,AS,Stngl}. 

\begin{figure}[htb]
\begin{center}
\includegraphics{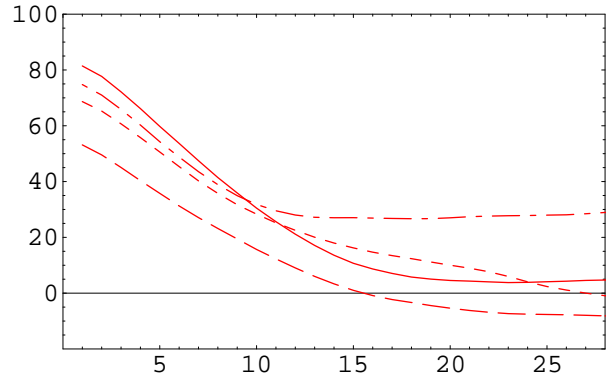}
\end{center}
\caption{The 1-d FT of the gluon propagator along the 4 axes. $\beta=6.4$,$56^4$ in the $\log U$ definition. sample $I_A$ \label{refpos29on}}
\end{figure}

\begin{figure}[htb]
\begin{center}
\includegraphics{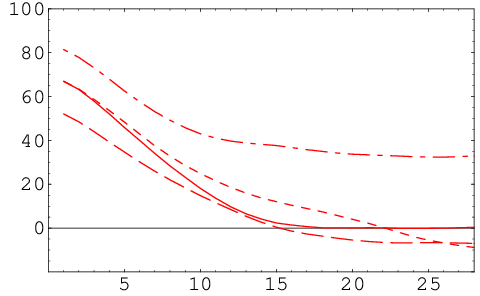}
\end{center}
\caption{The 1-d FT of the gluon propagator transverse to the 4 axes. $\beta=6.4$,$56^4$ in the $\log U$ dfinition. sample $I_B$ \label{refpos29nn}}
\end{figure}

The four 1-d FT of the copy $I_A$ and those of the copy $I_B$ are shown in Fig. \ref{refpos29on} and in
Fig. \ref{refpos29nn}, respectively. The solid line, dotted line, dashed line and the dash-dotted line corresponds to propagator transverse to $x_1,x_2,x_3$ and $x_4$ axis in the Euclidean space, respectively.

We observe that the gluon propagators of copies $I_A$ and $I_B$ have a specific axis along which the propagator manifestly violates reflection positivity. 
Here, manifestly means that it remains negative in a wide range in the intermediate not only in the large distance in the coordinate space. Propagators transverse to other axes 
in the copy $I_B$ are shifted from those of $I_A$ and the propagator almost parallel to that manifestly violating reflection positivity remains finite in the copy $I_A$, but it becomes almost 0 in the large distance in the copy $I_B$. The $L_2$ norm squared $\|A\|^2$ of copy $I_B$ is smaller than that of $I_A$, and hence $I_B$ is closer to the FMR but is not necessarily closer to the boundary of the Gribov region. Rather small shifts of the gluon propagators among copies make a significant difference in the exponent of the ghost propagator and the Kugo-Ojima parameter is surprising.

The ghost propagator is defined by the expectation value of the inverse Faddeev-Popov(FP) operator $\cal  M$
\begin{equation}
D_G^{ab}(x,y)=\langle  \langle \lambda^a x|({\cal  M}[U])^{-1}|
\lambda^b y\rangle \rangle,
\end{equation}
via the Fourier transform 
\begin{equation}\label{dgg}
D_G(q^2)=\frac{G(q^2)}{q^2}.
\end{equation}

The Kugo-Ojima parameter is defined by the two point function of the
covariant derivative of the ghost and the commutator of the antighost and gauge field
\begin{eqnarray}
&&(\delta_{\mu\nu}-{q_\mu q_\nu\over q^2})u^{ab}(q^2)\nonumber\\
&&={1\over V}
\sum_{x,y} e^{-ip(x-y)}\langle  {\rm tr}\left({\lambda^a}^{\dag}
D_\mu \displaystyle{1\over -\partial D}[A_\nu,\lambda^b] \right)_{xy}\rangle.
\end{eqnarray}

We performed the same analyses as sample $I$ for a sample which has the second largest Kugo-Ojima parameter (samples $II_A$ and $II_B$). 
The sample dependences of the $L_2$ norm of the gauge field, Kugo-Ojima parameter $c=-u(0)$, trace divided by the dimension $e/d$, horizon function deviation parameter $h$ \cite{Zw, NF} and the infrared exponent of the ghost propagator at 0.4GeV region $\alpha_G$, are summarized in Table \ref{gribovc}. Errors in $c$ is due to the deviation of the tensor sturucture from $(\delta_{\mu\nu}-{q_\mu q_\nu\over q^2})$ i.e. $c$ depends on the choice of $\mu$ as in the exceptional copy. 

 We parametrize infrared power dependence of $D_A(q^2)$ as $\simeq (qa)^{-2(1+\alpha_D)}$
and that of $D_G(q^2)$ as $\simeq (qa)^{-2(1+\alpha_G)}$.
Errors in $\alpha_G$ are estimated from the standard deviation in the plot of $\log D_G$ as a function of $\log q$ and we find they are $(-0.1,+0.45)$.  
An analysis of DSE \cite{FAR} suggests that the exponent at 0.4GeV is about half of the asymptotic value $\kappa$ and thus $\alpha_G$ corresponds to about half of $\kappa$. From our standard deviation of $\alpha_G$, we expect $\kappa$ in the range of $[0.1,0.7]$.
 
\begin{table}[hob]
\caption{The Gribov copy dependence of the Kugo-Ojima parameter $c$, trace divided by the dimension $e/d$, horizon condition deviation parameter $h$ and the 
exponent $\alpha_G$. }\label{gribovc}
\begin{center}
\begin{tabular}{c|cc|cc|c}
   &$I_A$ &$I_B$ & $II_A$ & $II_B$ & average \\
\hline
$\|A\|^2$ &0.09081&  0.09079 &   0.090698 & 0.090695 & 0.09072(7)\\
$c$ &0.851(77)&  0.837(58)  &  0.835(53) & 0.829(56) & 0.827(15)\\
$e/d$ &0.9535(1)&  0.9535(1)  &  0.9535(1) & 0.9535(1) & 0.954(1)\\
$h$ & -0.102(77) & -0.117(58) & -0.118(53) & -0.125(56) & -0.127(15) \\
$\alpha_G$ &0.272&  0.241  &  0.223 & 0.221 & 0.223\\
\hline
\end{tabular}
\end{center}
\end{table}

We observed that in most samples the dependence of the copy on $|\partial A|_{cr}$ is weak as in the case of sample $II$, and that the large difference of $I_A$ and $I_B$ copies is exceptional.  The Table \ref{gribovc} also shows that  $\alpha_G$, $c$ and $h$ are correlated. 
In the average of 15 samples of $56^4$ lattice data, we incorporate copy $A$ 
but not $B$.
 The  $\alpha_G$ of the sample average is 0.22, but that of the $I_A$ copy is 0.27.  The $I_A$ copy has a larger $L_2$ norm of the gauge field but smaller $h$ and larger $c$.  We find that not all samples have the axis that manifestly violates reflection positivity and that the direction of the axis is sample dependent.

\subsection{Kugo-Ojima parameter}

Our sample average of $c=-u(0)$, $e/d$, $h$, the exponent of the ghost dressing function $\alpha_G$, the exponents of the gluon dressing function $\alpha_D$ near $q=0.4GeV$, and $\alpha_D'$ near $q=1.97$GeV  are summarized in Table \ref{kugotab}.  

\begin{table}[htb]
\caption{The Kugo-Ojima parameter $c$, trace divided by the dimension $e/d$, horizon function deviation $h$ in  the  $\log U$ definitions.
The exponent of the ghost dressing function near zero momentum $\alpha_G$, the exponent of the gluon dressing function near zero momentum  $\alpha_D$, near $q=1.97$GeV  $\alpha_D'$ in  $\log U$ type. $\beta=6.0$ and $6.4$. }\label{kugotab}
\begin{center}
\begin{tabular}{c|ccc|ccc}
$\beta$ &      & 6.0     &          &          & 6.4      &     \\
\hline
$L$ & 16       & 24      & 32       & 32       & 48       & 56  \\
\hline
$c$ & 0.628(94)&0.774(76)& 0.777(46)& 0.700(42)& 0.793(61)&0.827(27)\\
$e/d$ & 0.943(1)&0.944(1)&0.944(1) &0.953(1)&0.954(1)&0.954(1) \\
$h$   & -0.32(9)   & -0.17(8) & -0.16(5) & -0.25(4) & -0.16(6) & -0.12(3) \\
$\alpha_G$  & 0.175   & 0.175  & 0.174   & 0.174  & 0.193 & 0.223 \\ 
$\alpha_D$ &     & -0.310 & -0.375  &    & -0.273 & -0.323 \\
$\alpha_D'$ & 0.38   & 0.314  & 0.302   & 0.31   & 0.288 & 0.275 \\
\hline
\end{tabular}
\end{center}
\end{table}

The color off-diagonal, space diagonal part of the Kugo-Ojima parameter $c$ was
0.0001(162) and consistent to 0.
The magnitude of the Kugo-Ojima parameter $c$ and exponent of the ghost propagator $\alpha_G$ are tightly correlated and they are also correlated with the violation of the reflection positivity in the gluon propagator. In the $I_A$ copy, reflection positivity is violated along $x_3$ axis and the average of 33 and 88 color components of $c$ along this axis is 0.97(6), consistent with 1. 

\subsection{Gluon propagator}
\begin{figure}[htb]
\begin{center}
\includegraphics{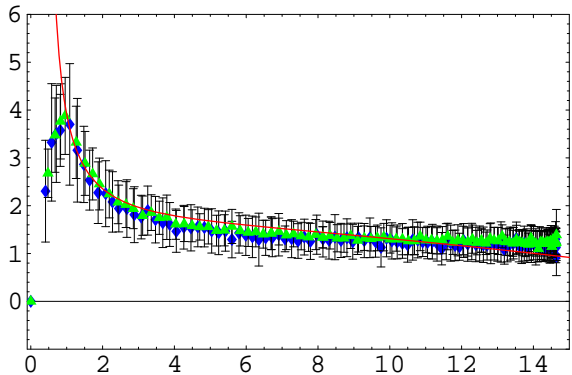}
\end{center}
\caption{The gluon dressing function as the function of the momentum $q$(GeV).  $\beta=6.4$, $48^4$(triangles) and $56^4$(diamonds) in the $\log U$ definition,
extrapolated to $V=\infty$. The solid line is that of the $\widetilde{\rm MOM}$ scheme. All data are scaled at $\mu=9.5GeV$.}\label{gl4856}
\end{figure}

The gluon propagator in momentum space was measured by using cylindrical cut method \cite{adelaide}, i.e., choosing momenta close to the diagonal direction.
In Fig. \ref{gl4856} we show the gluon dressing function of $\beta=6.4, 56^4$
lattice data together with $48^4$ lattice data.
The gluon propagators of $24^4, 32^4$ and $48^4$ as a function of the physical momentum agree with each other within errors and they can be fitted by the $\widetilde{MOM}$ scheme in two loop perturbation theory\cite{FN03,vAc}.
\begin{equation}
D_A(q^2)=\frac{Z(q^2,y)|_{y=0.02227}}{q^2}=\frac{Z_A(q^2)}{q^2}
\end{equation}
The overall normalization in this fitting turned out to be problematic since
the $56^4$ data are suppressed than the $48^4$ data.  We remove the lattice artefact by rescaling the data of the dressing function to that of the fit in the $\widetilde{\rm MOM}$ scheme $Z_A(9.5GeV)=1.3107(9)$ \cite{Orsay2}.


\subsection{Ghost propagator}

\begin{figure}[htb]
\begin{center}
\includegraphics{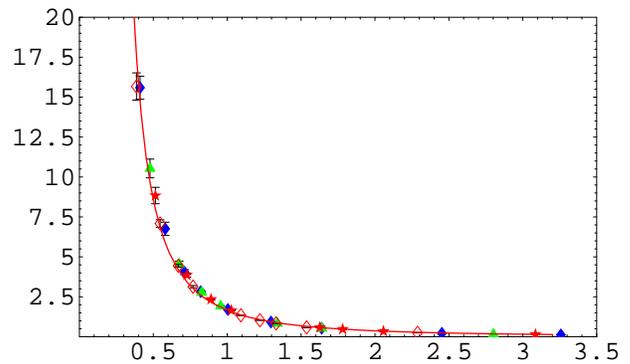}
\end{center}
\caption{The ghost propagator as the function of the momentum $q$(GeV). $\beta=6.0$, $24^4$(star),$32^4$(unfilled diamond),$\beta=6.4$, $48^4$(triangle) and $56^4$(filled diamond) in the $\log U$ definition. The fitted line is that of the $\widetilde{\rm MOM}$ scheme. }\label{gh4856}
\end{figure}

The ghost dressing function is defined by the ghost propagator as $G^{ab}(q^2)=q^2 {D_G}^{ab}(q^2)$. 
In Fig. \ref{gh4856}, $\beta=6.4$, $48^4$, and $56^4$ and
$\beta=6.0$ $24^4$ and $32^4$ lattice data of the ghost propagator are compared with that of the $\widetilde{\rm MOM}$ scheme\cite{vAc,FN03}. 
\begin{equation}\label{dg}
D_G(q^2)=-\frac{Z_g(q^2,y)|_{y=0.02142}}{q^2}=\frac{G(q^2)}{q^2}.
\end{equation}

We observe that the agreement is good for $q>0.5$ GeV. The $\widetilde{\rm MOM}$
 scheme is singular at $\tilde\Lambda_{\overline{\rm MS}}\simeq 0.35$ GeV but the singularity should be shifted to 0 momentum by the non-perturbative effects. The ghost propagator
was first measured in \cite{SuSh} but the scaling property was not observed and
the lowest momentum point was incorrectly suppressed. It may worth while to
remark that the rescaling is unnecesary in the ghost propagator of different lattice sizes, but the scale depends on the definition of the gauge field. The propagator of $\log U$ definition is about 14\% suppressed from that of the $U-$ linear definition. 

\subsection{QCD running coupling}

We measured the running coupling from the product of the gluon dressing function and the ghost dressing function squared \cite{SHA,BCLM1}. In terms of exponents $\alpha_D$ and $\alpha_G$, the running coupling near 0.4GeV is parametrized as
\begin{equation}
\alpha_s(q^2)=\frac{g_0^2}{4\pi}\frac{Z_A(q^2){ G(q^2)}^2}{{\tilde Z_1}^2}\simeq (qa)^{-2(\alpha_D+2\alpha_G)}.
\end{equation}

The lattice size dependences of the exponents $\alpha_D$ and $\alpha_G$ are summarized in Table \ref{kugotab}. 

The vertex renormalization factor $\tilde Z_1$ is 1 in the perturbation theory, but on the lattice it is not necessarilly the case. By comparing data of various $\beta$, finiteness of $\tilde Z_1$ was confirmed in the case of SU(2) \cite{BCLM}. 
 In the present analysis, we fix $\tilde Z_1$ by normalizating the running coupling by that of the perturbative QCD near the highest momentum point. 
In the lattice simulation of the three gluon coupling \cite{Orsay}, the nonperturbative
effect is found to be significant even at 10 GeV region, and a fit of the lattice data by the three loop perturbative term plus $c/q^2$ correction was proposed.
 We normalize the running coupling to that of Orsay group at the point of 14.4 GeV, i.e. 0.154(1).  This correction 
revises the previous results of $48^4$ lattice data\cite{FN03} by a factor of
1.97, and the maximum of the running coupling becomes 2.0(3).

\begin{figure}[htb]
\begin{center}
\includegraphics{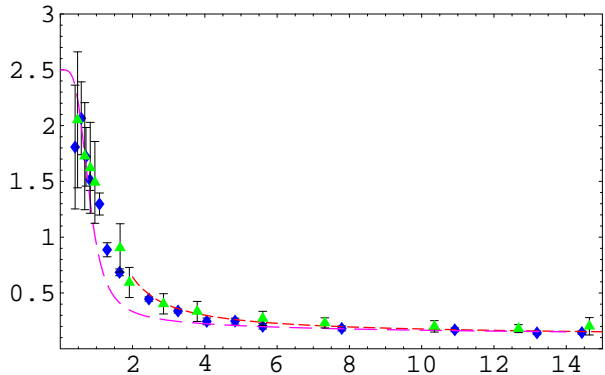}
\end{center}
\caption{The running coupling $\alpha_s(q)$ as a function of momentum $q$(GeV) of the $\beta=6.4$, $56^4$ lattice and $48^4$ lattice. The DSE approach with $\alpha_0=2.5$ (long dashed line) and the  the Orsay group(dotted line) are also plotted.
}\label{alp4856c}
\end{figure}

 In Fig.\ref{alp4856c} we present the rescaled running coupling of $48^4$ lattice  and that of the $56^4$ lattice and the fit of Orsay group above 2GeV and the result of the  MR truncation scheme of Bloch\cite{Blo,Blo1}, where in addition to the sunset diagram, the squint diagram was included. The running coupling in this DSE is parametrized as
\begin{eqnarray}\label{alpfit}
&&\alpha_s(q^2)=\alpha(t\Lambda_{QCD}^2)\nonumber\\
&&=\frac{1}{c_0 + t^2}(c_0 \alpha_0 +\frac{4\pi}{\beta_0}
(\frac{1}{\log t} -  \frac{1}{t - 1})t^2)
\end{eqnarray}
where $t=q^2/\Lambda_{QCD}^2$. The infrared fixed point $\alpha_0$ is expressed as an analytic function of $\kappa$, and \cite{Blo1} claims that when two-loop squint diagrams are included, possible solutions exist only for $\kappa$ in the range $[0.17,0.53]$.  Conjectures from DSE \cite{Blo1, Kond} predicts $\kappa\sim 0.5$, which implies $\alpha_0\sim 2.5$.

Except the value of the lowest momentum point, our data is consistent with the prediction $\alpha_0=2.5$. Thus, we 
adopt this value for $\alpha_0$ and search the parameter $c_0$ by the fit to the second and the third lowest momentum points of the running coupling. We find parameter $c_0=30$, instead of $c_0=15$ in the DSE \cite{Blo}.

Phenomenologically fitted $\Lambda_{QCD}$ from $\alpha(M_Z)$ is about 710 MeV, but the value depends on the number of quark flavors and in the quenched approximation the choice is not appropriate. We choose as \cite{Blo}, $\Lambda_{QCD}=330$ MeV.

\begin{figure}[htb]
\begin{center}
\includegraphics{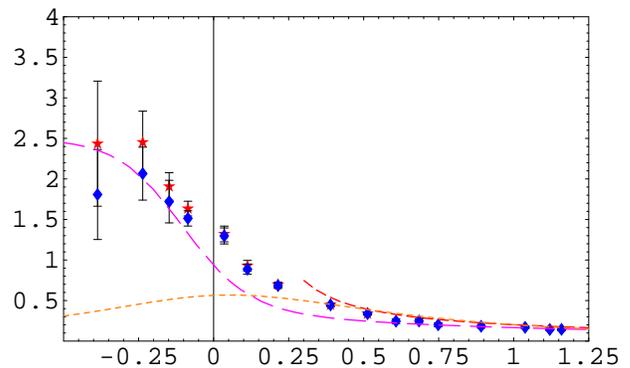}
\end{center}
\caption{The running coupling $\alpha_s(q)$ as a function of the logarithm of momentum $\log_{10}q$(GeV) of the $\beta=6.4$, $56^4$ lattice using the ghost propagator of the $I_A$ copy (stars) and that of the average (diamonds). The DSE approach with $\alpha_0=2.5$(long dashed line), the fit of the Orsay group perturbative +$c/q^2$ (short dashed line) and the contour improved perturbation method ( dotted line) are also shown.
}\label{alp56}
\end{figure}

When the ghost propagator of the exceptional copy is adopted, suppression of the running coupling at 0 momentum disappears. 
The DSE results, Orsay fit and the lattice data of the running coupling in which the ghost dressing function is taken from the average as a function of logarithm of momentum $\log_{10}q$(GeV) are shown in Fig.\ref{alp56}. 
In order to show the dependence on the Gribov copy, the data in which the ghost dressing function is replaced by that of the $I_A$ copy is also shown in the same figure. The ensemble of gluon propagator was not changed in this replacement, since the sample-wise difference of the gluon propagator is insignificant.

The contour improved perturbation method  with $\Lambda=e^{70/6\beta_0}\tilde\Lambda_{\overline{MS}}$ in two loop order \cite{HoMa, FN03} is consistent with our data at $q >10$GeV region, (dotted line) but in the infrared region it underestimates the lattice data. The dotted line is qualitatively the same as the
results of hypothetical $\tau$ lepton decay \cite{Bro}. 

\section{Comparison with DSE and ERGE}

 In the DSE approaches, infrared power behavior and specific relation between the exponent of the ghost propagator and the
gluon propagator is assumed. In the ERGE, flow equation in terms of the effective average action $\Gamma_\Lambda$ where $\Lambda$ is the infrared cut-off scale
is considered\cite{Wet,EHW,Kato}. In a recent work four point vertices in addition to the two point vertices are incorporated and the running coupling was calculated via
\begin{equation}
\alpha(q^2)=\frac{g^2(\Lambda_0)}{4\pi f_Z(q^2;\Lambda\to 0){f_G}^2(q^2;\Lambda\to 0)}
\end{equation}
where $f_Z(q^2;\Lambda)$ and $f_G(q^2;\Lambda)$ are gluon and ghost propagator function, respectively. They are related to the gluon and ghost propagator as
\begin{equation}
D_{A,\mu\nu}(q^2)=(\delta_{\mu\nu}-q_\mu q_\nu/q^2)\frac{1}{q^2 f_Z(q^2;\Lambda\to 0)}
\end{equation} 
and \begin{equation}
D_G(q^2)=-\frac{1}{q^2 f_G(q^2;\Lambda\to 0)}
\end{equation} 
The infrared exponent $\kappa$ obtained in this analysis turned out to be
$\kappa\sim 0.146$ in contrast to the DSE approach which suggested $\kappa\sim 0.5$.  The infrared fixed point $\alpha_0\sim 4.70$ was predicted\cite{Kato} which is about factor 2 larger than our lattice simulation. There is a prediction $\kappa=0.59535\cdots$ and the infrared fixed point $\alpha_0=2.9717\cdots$ both in DSE and ERGE\cite{LvS,vS}. 

 The prediction $\alpha_0=2.6$ and $\kappa=0.5$ of \cite{Blo1} is consistent with our lattice data. Here we summarize his approach and compare our lattice results.

The quark propagator in Euclidean momentum state is expressed as\cite{Blo, FA}
\begin{equation}
\frac{1}{-iq_\mu \gamma_\mu A(q^2)+B(q^2)}=\frac{Z(q^2)}{-iq_\mu \gamma_\mu+M(q^2)} 
\end{equation}
and $M(q^2)=B(q^2)/A(q^2)$ is proportional to the quark condensate at large $q^2$:
\begin{equation}
M(q^2)\sim m_\mu-\frac{4\pi\alpha_s(q^2)}{3q^2}(\frac{\alpha_s(q^2)}{\alpha_s(\mu^2)})^{-d_m}\langle \bar \psi \psi(\mu^2)\rangle
\end{equation}
where $d_m=12/(33-2N_f)$. Here the number of flavor $N_f=0$ in the quenched approximation.

The quark field is renormalized as
\begin{equation}
Z(q^2,\mu^2)=Z_2(\mu^2,\Lambda^2)Z_R(q^2,\mu^2)
\end{equation}
where $Z_R$ is the renormalized quark dressing function, $Z_2$ is the quark field renormalization constant and at the renormalization point. We define $Z_R(x)=Z_R(x,\mu^2)$ and $Z_R(\mu^2)=1$ and $m_\mu=M(\mu^2)$. In the DSE\cite{Blo1}, $\mu$ is chosen to be $\Lambda_{QCD}=330$MeV. 

The renormalized quark dressing function $Z_R(q)$  and the quark mass function $M(q)$ can be calculated 
by a coupled equation once the running coupling $\alpha_s(q^2)$ is given \cite{Blo}. 
The quark mass function at the origin $M(0)$ is a function of the parameter $c_0$ and our fitted value $c_0=30$ yields
\begin{equation}
M(0)\simeq 1.27\Lambda_{QCD}=0.419 {\rm GeV}.
\end{equation}
This value is consistent with the result of quark propagator in quenched lattice Landau gauge simulation\cite{BBLWZ} extrapolated to $0$ momentum. 
The quark condensate $\langle \bar \psi\psi\rangle$ is estimated as $-(0.70\Lambda_{QCD})^3$ which is compatible with the recent analysis of quenched lattice QCD \cite{BL}. 

When $\kappa$ is larger than 0.5 as predicted by \cite{LvS,Zwa1}, the gluon
propagator should vanish in the infrared. The present lattice data are not
compatible with this prediction.

\section{SU(2) $16^4$ lattice data}
In \cite{BCLM1}, finiteness of the vertex renormalization factor $\tilde Z_1$ was prooved by linear rising of the $Z_A(\mu^2)G(\mu^2)^2$ ($\mu=3$GeV) as a function of $-\log(a(\beta)^2 \sigma)$ where $a(\beta)$ is the lattice spacing corresponding to the $\beta$ and $\sigma=[440MeV]^2$ is the string tension.   In order to check this behavior and to see infrared features of the SU(2) lattice Landau gauge, we performed Monte Carlo simulation of SU(2) lattice Landau gauge using
the $U-$linear definition of the gauge field. We choose $\beta=2.2, 2.3$ and 2.375 and accumulate 200 samples for each $\beta$.

We confirmed increasing of $Z_A(\mu^2)G(\mu^2)^2$ ($\mu=3$GeV) from $\beta=2.3$
to $2.375$ with the slope $\gamma$ consistent with $13/22$. The data of $\beta=2.2$ was off the fitted line, but we expect this is due to the closeness of the $\mu=3GeV$ point to the maximal momentum point 3.7GeV.

\begin{figure}[htb]
\begin{center}
\includegraphics{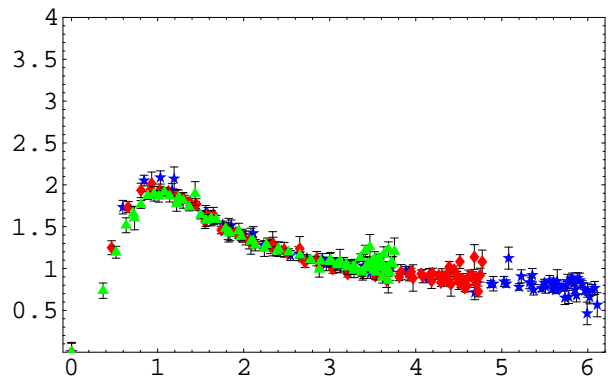}
\end{center}
\caption{The SU(2) gluon dressing function as a function of momentum $q$(GeV) of the $\beta=2.2$(triangles), 2.3(diamonds) and 2.375(stars), $16^4$ lattice (200 samples). 
}\label{gldsu2nn}
\end{figure}

\begin{figure}[htb]
\begin{center}
\includegraphics{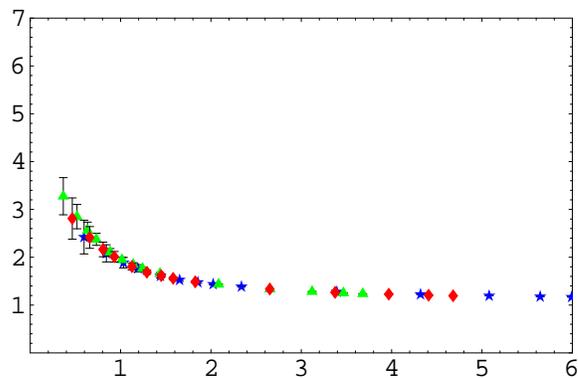}
\end{center}
\caption{The SU(2) ghost dressing function as a function of momentum $q$(GeV) of the $\beta=2.2$(triangles), 2.3(diamonds) and 2.375(stars), $16^4$ lattice (200 samples). 
}\label{ghdsu2nn}
\end{figure}

The gluon dressing function and the ghost dressing function as a function of the momentum $q$(GeV) of the $\beta=2.2$, 2.3 and 2.375 are shown in Fig.\ref{gldsu2nn} and in Fig.\ref{ghdsu2nn}, respectively.  In the gluon dressing function, cylindrical cut is applied and the error bars are obtained by the jacknife method. Error bars of the ghost dressing function are the standard deviation.

\begin{figure}[htb]
\begin{center}
\includegraphics{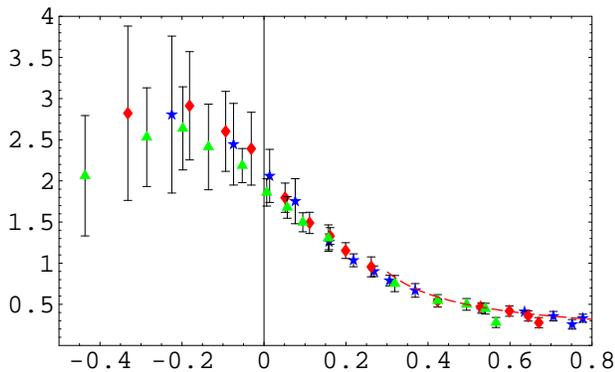}
\end{center}
\caption{The SU(2) running coupling $\alpha_s(q)$ as a function of the logarithm of the momentum $\log_{10} [q$(GeV)] of the $\beta=2.2$(triangles), 2.3(diamonds) and 2.375(stars), $16^4$ lattice (200 samples). The result of two-loop perturbation theory (dotted line ) are also plotted. 
}\label{alpsu2nn}
\end{figure}

The running coupling $\alpha_s(q)$ as a function of the logarithm of the momentum $\log_{10} [q$(GeV)] of $\beta=2.2,2.3$ and 2.375 are plotted in Fig.\ref{alpsu2nn}. 
We normalize the running coupling near the highest momemtum point by that of the two-loop perturbation results. This correction revises the previous result of the running coupling of SU(2) \cite{FN03} by about factor 1.54, but there remains  difference from Tuebingen and S\~ao Carlos \cite{BCLM1} by about factor 2. 

The ghost propagator and the gluon propagator of \cite{BCLM1} were rescaled 
by the tadpole renormalization factor $u_P$. There are qualitative agreements in ghost propagator of \cite{BCLM1} and ours, but in the gluon propagator there are discrepancies in the momentum dependence in the infrared region.  In \cite{adelaide1} it is remarked that in simulations of relatively small lattice with a lattice axis chosen to be twice as those of the other three lattices, the gluon propagator of a few lowest momentum points do not match smoothly to those of higher momenta. Our gluon propagator below 1 GeV is more suppressed than those of Tuebingen data of $16^3\times 32$\cite{BCLM1}, and the discrepancy could be due to this finite size effect. Tuebingen group adopts adjoint links and thus their tadpole renormalization makes direct comparison of the gluon propagators obscure. In the running coupling, however, the tadpole renormalization factor $u_P$ for the ghost and for the gluon cancel \cite{BCLM1}, and the difference in the running coupling $\alpha_s(q)$ in the infrared is due to the difference in the shape of the gluon propagator. 

\section{Unquenched SU(3) $20^3\times 48$ lattice data}
We observed that there are samples whose 1-d FT of the gluon propagator transverse to a lattice axis manifestly violates reflection positivity.
The direction of the reflection positivity violating axis appears randomly. 
 Recently, Aubin and Ogilvie \cite{AO} pointed out that the origin of the reflection positivity violation lies in the quenched character of the gauge transformation $g$. They demonstrared in a Higgs model type  SU(2) $20^4$ lattice simulation, occurlence of reflection positivity violation analogous to that in the quenched lattice simulation of $a_0$ meson propagator, by considering the gauge transformation of $G_{local}\times G_{global}$. In order to see qualitative difference between quenched and unquanched simulation, and to investigate finite size effects in lattices whose one axis is taken longer than the others, we studied infrared features of the unquenched SU(3) $20^3\times 48$ lattice configuration of the JLQCD\cite{jlqcd}, where improved Wilson action with Sheikholeslami and Wohlert parameter $c_{SW}=2.02$ and the number of sea quark flavours $N_f=2$ are adopted. We choose SHMC algorithm configurations of hopping parameter $K_{sea}=0.1340$ and 0.1355.

We performed the Landau gauge fixing on 9 samples for each $K_{sea}$ using the $\log U$ definition for the gauge field and measured the gluon propagator, ghost propagator, QCD running coupling
and the Kugo-Ojima parameter.  In addition to the correlation of gauge fields around the diagonal $[q_1,q_2,q_3,q_4]=[q,q,q,(48 q/20)]$ where $(48 q/20)$ is an integer close to this quotient, we measured the correlation transverse to the
coordinate axis $x_i$ as
\begin{eqnarray}
D_{A,kl}(q)&=&{1\over n^2-1}\sum_{x={\bf x},t}e^{-iqx}Tr\langle A_k(x)A_l(0)^\dagger \rangle \nonumber\\
&=&(\delta_{kl}-{q_k q_l\over q^2})D_A(q^2).
\end{eqnarray}
where $k$ and $l$ run over 1,2 and 3 $\ne i$, and the same expression for the time axis $x_4$. 
The four 1d-FT of these sample-wise gluon propagators turned out to be quite different from those of the quenched simulation.  As shown in Figs.\ref{refpos29on} and \ref{refpos29nn}, in quenched case in general, there is a component which remains positive in the whole region and different from other three components, but such a component is absent in the unquenched case in general. Although symmetry violation in each sample does not mean symmetry violation in the ensemble, the difference suggests that the global symmetry, i.e. rotational symmetry is recovered by the coupling of the gluon to fermions \cite{gosh}. 

\begin{figure}[htb]
\begin{center}
\includegraphics{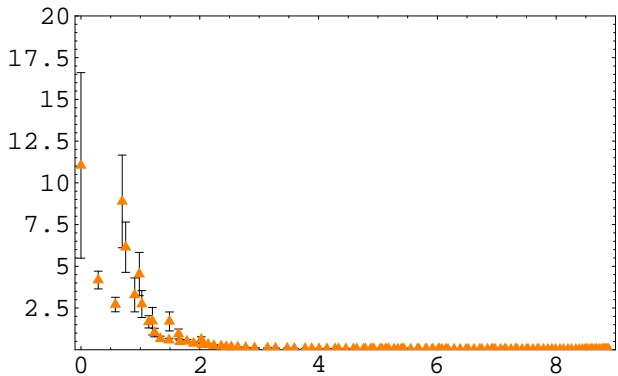}
\end{center}
\caption{The gluon propagator $D_A(q)$ as a function of the momentum $q$(GeV) of the $\beta=5.2$, $20^3\times 48$ lattice using the configuration of $K_{sea}=$0.1355. 
}\label{gl1355}
\end{figure}

When the length of an axis is taken longer than the other three axes, the
 $Z(4)$ symmetry corresponding to interchange of the axes is broken to $Z(3)$ symmetry and it is serious in the estimation of the infrared gluon propagator.
 In the
$20^3\times 48$ SU(3) unquenched lattice simulation,  the gluon propagator of the 3 lowest momentum
points ($p_1,p_2,p_3,p_4)=$(0,0,0,0),(0,0,0,1) and (0,0,0,2) do not match smoothly to higher momenta (Fig.\ref{gl1355}).  

The problem due to lack of rotational symmetry in the gluon propagator is
usually evaded by performing the cylindrical cut. In this context, the momentum
points (1,0,0,0) and its $Z(3)$ partners are farther than the (0,0,0,2) to the cylidrical axis and the treatment of these points remains a problem. In a preliminary calculation of the running coupling using (1,0,0,0) and its $Z(3)$ partners of the unquenched $20^3\times 48$ lattice, we found $\alpha_s$(0.98GeV)$\sim 3$ and the infrared fixed point of $\alpha_0\sim 4$ is suggested. The gluon propagator obtained by Landau gauge fixing unquenched SU(3) configuration in which L\"uscher-Weisz improved action is used shows that the rotational symmetry of $20^3\times 64$ lattice is recovered\cite{BHLPW}.
 Details of the investigation of the running coupling of unquenched SU(3) Landai gauge simulation will be presented elsewhere. 

\section{Discussion and outlook}
We measured the gluon dressing function and the ghost dressing function in lattice Landau gauge QCD and calculated the running coupling. In view of uncertainty
in the vertex renormalization factor $\tilde Z_1$ which is not necessarily 1 in the lattice simulation, we normalized the running coupling by that of perturbative QCD near the highest momentum point of the lattice. We found infrared fixed point $\alpha_0\sim 2.5(5)$, which is consistent with the MR scheme DSE calculation \cite{Blo1}. In the momentum dependence, there is disagreement with DSE in $2<q<10$ GeV region, which suggests a correction like $c/q^2$ term in $\alpha_s(q)$ \cite{Bouc}. Although this correction applies only in $q > 2$ GeV region, it could yield attraction between colored sources. 

 We observed that the 1-d FT of gluon propagator of the $I_A$ copy has an axis along which the reflection positivity is manifestly violated. The average of Cartan subalgebra components of Kugo-Ojima parameter along this specific axis becomes consistent with $c=1$. 
The 1-d FT of the gluon propagator transverse to the diagonal direction in the lattice is also performed by using the analytical expression of the gluon dressing function in $\widetilde{MOM}$ scheme for $q > 1$ GeV and numerical interpolation for $0< q <1$ GeV. In this ensemble average, violation of reflection positivity is very weak, although the quantitative feature is sensitive to the dressing function near $q=0$. 

When the QCD running coupling in the infrared region is thought to be divergent,
the dynamical chiral symmetry breaking was thought to be irrelevant to confinement\cite{Hig}. Our lattice data of running
coupling is qualitatively similar to that assumed in the model of dynamical chiral symmetry breaking. 

In passing, we compare running coupling measured in other lattice simulations.
Orsay group measured the running coupling with use of $U-$linear definition and from triple gluon vertex. The running coupling turned out to behave as $\propto p^4$ in the infrared contrary to ours, but above 0.8 GeV the data are consistent with ours. They
analyzed the infrared behavior in the instanton liquid model \cite{Orsay}. Running coupling around 0.2 GeV in instanton scheme using $U-$ linear definition measured by the DESY group \cite{RS} is $\alpha_s=4\sim 5$. 
A comparison of the $\log U$ and $U-$linear definitions of $48^4$ and $56^4$ are presented in \cite{NF,NF04}.
 The ghost propagator in $U-$linear definition is larger than that of $\log U$ definition, but the dependence does not explain the discrepancies in the running coupling from the DESY data, and we suspect problems in  finite size effects due to asymmetric shape of the lattice. 
 
In the study of instantons, Nahm conjectured that Gribov copies cannot tell much about confinement \cite{Nah}. We showed that the dynamical chiral symmetry breaking and confinement can be explained by using the same running coupling  and that the Gribov copy gives information on the ambiguity in the parameter that characterizes chiral symmetry breaking and confinement. We are currently analyzing infrared properties of the unquenched JLQCD configurations, i.e. the quark mass dependence of the Kugo-Ojima parameter and the running coupling. The results will be published in the future.

The running coupling of the quenched SU(3) simulation of our Landau gauge fixing suggests that there is a peak of $\alpha_s\sim 2.2$ at $q\sim 0.5$GeV, but the
running coupling calculated by the ghost propagator of the exceptional sample is consistent with the result of DSE with infrared fixed point $\alpha_s(0)\sim 2.5$.  Whether the population of the exceptional configuration  becomes larger when the system approaches to the continuum limit will be investigated.

\begin{acknowledgments}
We thank the referees for suggesting normalization of the running coupling
and the gluon dressing function by those of perturbative QCD in high momentum region.
S.F. thanks Kei-Ichi Kondo for attracing our attention to the ERGE approach.
Thanks are also due to the JLQCD collaboration for providing us their unquenched SU(3) configurations. 
This work is supported by the KEK supercomputing project No. 03-94. 
\end{acknowledgments}

\end{document}